\documentclass[12pt]{article}
\usepackage{amsmath,amsfonts,epsf}
\usepackage{amssymb}
\usepackage{graphicx}
\usepackage{grffile}
\input epsf

\makeatletter\@addtoreset{equation}{section}\makeatother

\def\be{\begin{equation}}
\def\ee{\end{equation}}
\def\bea{\begin{eqnarray}}
\def\eea{\end{eqnarray}}
\def\ie{\begin{equation}\begin{aligned}}
\def\fe{\end{aligned}\end{equation}}

\newcommand{\A}{{\alpha}}
\newcommand{\B}{{\beta}}
\newcommand{\C}{{\gamma}}
\newcommand{\D}{{\delta}}
\newcommand{\da}{{\dot\alpha}}
\newcommand{\db}{{\dot\beta}}
\newcommand{\dc}{{\dot\gamma}}
\newcommand{\dd}{{\dot\delta}}

\makeatletter\@addtoreset{equation}{section}\makeatother

\renewcommand{\title}[1]{\vbox{\center\LARGE{#1}}\vspace{5mm}}
\renewcommand{\author}[1]{\vbox{\center#1}\vspace{5mm}}
\newcommand{\address}[1]{\vbox{\center\em#1}}
\newcommand{\email}[1]{\vbox{\center\tt#1}\vspace{5mm}}

\begin{document}
\begin{titlepage}
\begin{flushright}
PI-strings-222
\end{flushright}
\begin{center}
\hfill \\
\hfill \\
\vskip 1cm

\title{A Note on CFT Correlators in Three Dimensions}


\author{Simone Giombi$^{1,a}$, Shiroman Prakash$^{2,b}$ and
Xi Yin$^{3,c}$}

\address{${}^1$Perimeter Institute for Theoretical Physics, Waterloo, Ontario, N2L 2Y5, Canada}
\address{
${}^2$Department of Theoretical Physics,
Tata Institute of Fundamental Research,\\
Homi Bhabha Road,
Mumbai 400005, India}
\address{
${}^3$Center for the Fundamental Laws of Nature,
Jefferson Physical Laboratory,\\
Harvard University,
Cambridge, MA 02138 USA}

\email{$^a$sgiombi@pitp.ca, $^b$shiroman@gmail.com,
$^c$xiyin@fas.harvard.edu}

\end{center}

\abstract{In this note we present a simple method of constructing general conformally invariant three point functions of operators of various spins in three dimensions. Upon further imposing current conservation conditions, we find new parity violating structures for the three point functions involving either the stress-energy tensor, spin one currents, or higher spin currents. We find that all parity preserving structures for conformally invariant three point functions of higher spin conserved currents can be realized by free fields, whereas there is at most one parity violating structure for three point functions for each set of spins, which is not realized by free fields. }

\vfill

\end{titlepage}


\section{Introduction}

The holographic duality between quantum theories of gravity in $AdS_4$ and three dimensional CFTs \cite{Maldacena:1997re} is particularly rich and intriguing for several reasons. Firstly, it provides in principle a complete description of quantum gravity in four dimensions. Such a description could be used to address the existence of UV completions of pure gravity or gravity coupled to a few matter fields in four dimensions, as well as to provide a computable framework for gravity interactions at (super-)Planckian scale. Secondly, three dimensional CFTs are extraordinarily rich. In particular, there is a large family of 3d CFTs with the Lagrangian description in terms of Chern-Simons-matter (CSM) theories \cite{Ivanov:1991fn}. Generally, these theories can be studied perturbatively. In some cases, exact nonperturbative results are also possible (such as theories with vector-like matter in the large $N$ limit, or certain BPS observables in supersymmetric theories that can be computed via localization method). Thirdly, there are conjectures of {\sl exact} dualities between nonsupersymmetric vector-type 3d CFTs and higher spin gravity theories in $AdS_4$ \cite{Sezgin:2002rt, Klebanov:2002ja}. Strong evidences in support of these dualities have been found recently \cite{Giombi:2009wh, Giombi:2010vg}, and a computable framework has been formulated which allows one to compare the boundary and bulk order by order in the $1/N$ expansion.

Thus far, the AdS/CFT correspondence has been mostly useful only when the boundary theory has a $1/N$ expansion, corresponding to the perturbative expansion in the bulk gravity coupling. A ``typical" large $N$ CFT has primary operators of various spins and dimensions that do not scale with $N$. These operators are dual to various higher spin fields in the bulk.
In this note, we will explore constraints on correlation functions of higher spin operators in a 3d CFT, from conformal symmetry alone, extending earlier work of Osborn and Petkou \cite{Osborn:1993cr}. Although we do not expect interacting 3d CFTs to have exact higher spin symmetries, sometimes there can be enhanced higher spin symmetries in the infinite $N$ limit, which implies that certain correlators obey Ward identities for conserved currents at the leading nontrivial order in the $1/N$ expansion. We will consider both conserved and non-conserved higher spin currents, as well as spin one and spin two currents. We will focus on the three point functions. In particular, in the case of spin 1 and spin 2 conserved currents (denote by $j_\mu$ and $T_{\mu\nu}$), we find new, parity violating tensor structures for the three point functions $\langle Tjj\rangle$ and $\langle TTT\rangle$, which were not given in the previous work of Osborn and Petkou \cite{Osborn:1993cr} because parity has been implicitly assumed there. It will be demonstrated in \cite{GMPSY} that these parity odd structures are indeed present in parity violating Chern-Simons-matter theories. Parity violating structures are found for higher spin correlators as well.

In the next section, we describe the general conformally invariant three point functions without imposing current conservation. We then present a conjectural classification of all conformally invariant three point functions of conserved higher spin currents. Our main results are summarized in section 3.1. Various examples and tests are presented in subsequent sections. We conclude with some comments on the holographic duals.

\bigskip

Note: Upon completion of the bulk of this work, we became aware of \cite{Juan} which also observed the parity violating $\langle TTT\rangle$ structure.

\section{Correlators of higher spin operators}

\subsection{Constraints from conformal symmetry}

Denote by $J_{\mu_1\cdots\mu_s}(x)$ a primary operator of spin $s$. In particular, $J_{\mu_1\cdots\mu_s}$ is symmetric and traceless in $(\mu_1,\cdots,\mu_s)$. By using the Pauli matrices, we can map any such tensor to a $2s$-components multispinor  $J_{\A_1\cdots\A_{2s}} = J_{\mu_1\cdots\mu_s} \sigma^{\mu_1}_{\A_1\A_2}\cdots \sigma^{\mu_s}_{\A_{2s-1}\A_{2s}}$, which is completely symmetric in $(\A_1,\cdots,\A_{2s})$. 

It is convenient to write the higher spin current in terms of a generating polynomial, by contracting with a polarization spinor $\lambda^\A$,
\ie
J_s(x,\lambda) = J_{\A_1\cdots \A_{2s}}(x)\lambda^{\A_1}\cdots \lambda^{\A_{2s}}.
\fe
This is equivalent to contracting all the indices of $J_{\mu_1\cdots\mu_s}$ with a null polarization vector $\varepsilon^{\mu}$, related to $\lambda$ by
$\varepsilon_{\A\B}\equiv \varepsilon_\mu \sigma^\mu_{\A\B} = \lambda_\A \lambda_\B$. 

Consider the $n$-point function of primary operators of various spins,
\ie
F(\{x_i,\lambda_i, s_i\})=\left\langle \prod_{i=1}^n J_{s_i}(x_i,\lambda_i)\right\rangle\,.
\fe
In this section we will not assume that the currents $J_{s_i}(x_i)$ are conserved, but just that they are primary operators of conformal dimension $\Delta_i$, and we will only use constraints from conformal symmetry. It will turn out that the constraints from conformal symmetry are easy to implement in our approach, while the constraints from current conservation are more complicated. The consequence of imposing current conservation will be explored in Section 3.

The conformal group is generated by the Poincar\'e group together with dilatation $D$ and special conformal generators $K_\mu$. $K_\mu$ can be realized as $RP_\mu R$, where $R$ is the inversion $x^\mu\mapsto x^\mu/x^2$. Hence we look for Poincar\'e invariant structures with fixed weight under $D$ and invariant under $R$ up to a sign. Invariance under $D$ implies that $F(\{x_i,\lambda_i\})$ is homogeneous under the rescaling
\ie
x_i \mapsto t x_i, ~~~\lambda_i \mapsto t^{1\over 2} \lambda_i,
\fe
of degree $-\sum_i(\Delta_i-s_i)$. Under the inversion $R$, we have
\ie
x^\mu \mapsto \check x^\mu \equiv {x^\mu\over x^2},~~~ \lambda\mapsto {\bf \check x} \lambda \equiv \check\lambda,
\label{R-lam}
\fe
where ${\bf x}\equiv \vec x\cdot\vec \sigma$, ${\bf\check x} = {\bf x}/x^2$. The action by $R$ on $\lambda$ is defined such that $\varepsilon^\mu=\lambda\sigma^\mu\lambda$ transforms as
\ie
\varepsilon^\mu \mapsto \check\varepsilon^\mu = {\partial \check x^\mu\over\partial x^\nu} \varepsilon^\nu = {\varepsilon^\mu\over x^2} - {2\varepsilon\cdot x x^\mu\over x^4}.
\label{R-ep}
\fe
It follows that if $J_{\alpha_1\ldots \alpha_{2s}}(x)$ is a primary operator of spin $s$ and dimension $\Delta$, then the operator $J_s(x,\lambda)$ with all indices contracted with the auxiliary polarization spinor transforms under inversion as\footnote{The transformation rule of a spin $s$ primary of dimension $\Delta$ is $\check{J}_{\mu_1\cdots\mu_s}(\check x) = \omega(x)^{\Delta-s}\frac{\partial x^{\nu_1}}{\partial\check{x}^{\mu_1}}\cdots \frac{\partial x^{\nu_s}}{\partial\check{x}^{\mu_s}} J_{\nu_1\cdots \nu_s}(x)$, where $\omega(x)=|{\rm det \frac{\partial \check x}{\partial x}}|^{-1/3}$. For the inversion, $\omega(x)=x^2$. Therefore it follows from (\ref{R-ep}) that $J_s(x,\epsilon)=J_s(x,\lambda)$ transforms as in (\ref{R-J}).}
\ie
J_s(x,\lambda)\to \check J_s(\check x,\check\lambda)= \left(x^2\right)^{\Delta-s} J_s(x,\lambda),
\label{R-J}
\fe
namely, as a scalar primary of dimension $\Delta-s$. Invariance of the correlation function under $R$ then implies
\ie
F(\{\check x_i,\check\lambda_i, s_i\}) = \left(\prod_{i=1}^n \left(x_i^2\right)^{\Delta_i-s_i}\right) F(\{x_i,\lambda_i,s_i\})\,.
\fe
For given spins $s_i$, a priori, $F$ depends on $5n$ variables $(x_i,\lambda_i)$. The conformal group in three dimensions, $SO(3,2)$, has 10 generators. We thus expect to express $F$ in terms of $5n-10$ variables. From now on, we will focus on the $n=3$ case, where $F$ depends on $5$ independent variables, which should be expressed as Poincar\'e invariant combinations of $\{x_i, \lambda_i\}$.

Note that under $R$, ${\bf x}_{ij}$ and $x_{ij}^2$ transform as
\ie
& {\bf x}_{ij} \to {{\bf x}_i\over x_i^2}-{{\bf x}_j\over x_j^2} = -{\bf\check x}_i {\bf x}_{ij} {\bf\check x}_j,
\\
&x_{ij}^2 \to x_{ij}'^2 = {x_{ij}^2\over x_i^2 x_j^2},
\\
& {\bf\check x}_{ij} \to - {\bf x}_i{\bf\check x}_{ij} {\bf x}_j.
\label{R-bfx}
\fe
So expressions of the form
\ie
\lambda_{i_1}{\bf\check x}_{i_1i_2} {\bf x}_{i_2i_3}\cdots {\bf \check x}_{i_{n-1}i_n}\lambda_{i_n}
\fe
are invariant under the action of $R$ (note that $({\bf x}\lambda)^T = -\lambda^T {\bf x}$, and we have omitted the transpose in writing $\lambda\cdots\lambda$ with the understanding that the spinor indices are contracted from upper left to lower right). Therefore we see that the following six Poincar\'e invariant objects
\ie
& P_3 = \lambda_1 {\bf\check x}_{12}\lambda_2,~~ P_1 = \lambda_2 {\bf\check x}_{23}\lambda_3, ~~
P_2 = \lambda_3 {\bf\check x}_{31}\lambda_1, \\
& Q_1= \lambda_1 {\bf\check x}_{12}{\bf x}_{23}{\bf\check x}_{31}\lambda_1, ~~
Q_2=\lambda_2 {\bf\check x}_{23}{\bf x}_{31}{\bf\check x}_{12}\lambda_2, ~~
Q_3=\lambda_3 {\bf\check x}_{31}{\bf x}_{12}{\bf\check x}_{23}\lambda_3.
\label{even-structures}
\fe
are invariant under inversion, hence conformally invariant. We can alternatively write the $Q_i$'s as
\ie
Q_1 = - \lambda_1 ({\bf\check x}_{12}+{\bf\check x}_{31})\lambda_1,~~~Q_2 = - \lambda_2 ({\bf\check x}_{23}+{\bf\check x}_{12})\lambda_2,~~~Q_3 = - \lambda_3 ({\bf\check x}_{31}+{\bf\check x}_{23})\lambda_3.
\fe
Translating to vector notations, these objects correspond to the following well-known conformally invariant tensor structures \cite{Osborn:1993cr}
\ie
 (P_3)^2=-2\varepsilon_1^{\mu} \varepsilon_2^{\nu}\left[\frac{\delta^{\mu\nu}}{x_{12}^2}-2 \frac{x_{12}^{\mu}x_{12}^{\nu}}{x_{12}^4}\right]\,,\qquad
 Q_1 = 2 \varepsilon_1^{\mu}\left[\frac{x_{13}^{\mu}}{x_{13}^2}-\frac{x_{12}^{\mu}}{x_{12}^2}\right]\,,
\fe
and similarly for the remaining structures related to these by cyclic permutations. In particular $(P_3)^2$ is the structure appearing in the 2-point function of spin $s$ operators
\begin{equation}
\langle J_s(x_1,\lambda_1) J_s(x_2,\lambda_2) \rangle = c \frac{(P_3)^{2s}}{|x_{12}|^{2(\Delta-s)} }\,,
\label{JJ-2pt}
\end{equation} 
where $\Delta$ is the dimension of $J_s$, and $c$ some normalization constant. When $J_s$ is a conserved current, then it follows from the conformal algebra that $\Delta=s+1$. One can check that precisely for this value of $\Delta$ the above 2-point function indeed satisfies conservation. In $d=3$, conformal invariance also allows a parity odd contact term in the 2-point function of a spin $s$ current, see e.g. \cite{Leigh:2003ez}. In this paper, we study correlation functions in coordinate space at distinct points, and we will not keep track of contact terms.

Since the structures (\ref{even-structures}) are conformally invariant, any polynomial of $\{P_i, Q_i\}$ that is homogeneous of degree $2s_i$ with respect to $\lambda_i$, when multiplied by the scalar 3-point function factor $1/(|x_{12}|^{\delta_1+\delta_2-\delta_3}|x_{23}|^{\delta_2+\delta_3-\delta_1}|x_{31}|^{\delta_3+\delta_1-\delta_2})$,
with $\delta_i =\Delta_i-s_i$,  gives a tensor structure for $\langle J_{s_1}J_{s_2}J_{s_3}\rangle$ that is allowed by conformal symmetry.

Note that these $P_i, Q_i$'s are precisely the building blocks of the generating function of 3-point functions of higher spin currents in the free massless scalar and free massless fermion theory, which are respectively \cite{Giombi:2010vg}
\ie
& {\cal F}^{\rm bose}_{\langle JJJ \rangle} = \frac{1}{|x_{12}||x_{23}||x_{31}|} \cosh\left[\frac{1}{2}(Q_1+Q_2+Q_3)\right]\, \cosh P_1\, \cosh P_2\, \cosh P_3 \\
&{\cal F}^{\rm fermi}_{\langle JJJ \rangle} = \frac{1}{|x_{12}||x_{23}||x_{31}|} \sinh\left[\frac{1}{2}(Q_1+Q_2+Q_3)\right]\, \sinh P_1\, \sinh P_2\, \sinh P_3\,.
\label{free-gen}
\fe
The LHS are generating functions of $\langle J_{s_1} J_{s_2} J_{s_3}\rangle$ summed over the spins. In these free theories, the currents are of course conserved and $\Delta_i=s_i+1$. The free scalar expression is valid for $s_i \ge 0$ (the spin 0 ``current" is just the dimension 1 operator $\bar\phi \phi$), while the free fermion answer is valid for $s_i\ge 1$ (see Section \ref{3pt-scalar} below for the form of the generating function involving the $\Delta=2$ scalar operator $\bar \psi \psi$).

As remarked earlier, we expect that the 3-point functions should depend on $15-10=5$ independent conformally invariant objects.
Indeed, it can be seen that the six structures $(P_i,Q_i)$ satisfy the following non-linear relation
\begin{equation}
P_1^2 Q_1 + P_2^2 Q_2 + P_3^2 Q_3 - 2 P_1 P_2 P_3 - Q_1 Q_2 Q_3=0\,,
\label{cubic}
\end{equation}
thus leaving 5 independent degrees of freedom.

In theories with a parity symmetry, $J_{\mu_1\cdots\mu_s}$ acquires a sign $(-)^s$ under parity. We will assign $\lambda$ to transform as $\lambda\to i\lambda$ under parity, so that $J(x,\lambda)$ would be parity invariant.
The $P_i, Q_i$ structures defined above are even under the action of parity $x_k^{\mu}\rightarrow -x_k^{\mu}$, $\lambda_k\rightarrow i \lambda_k$, for $k=1,2,3$. In the spinor formalism, it is not difficult to see that there are in fact further independent conformal invariant structures built from bilinears of $\lambda_i$, which are odd under parity. Indeed, consider the structure 
\begin{equation}
\frac{\lambda_1 {\bf x}_{12} {\bf x}_{23}\lambda_3}{|x_{12}||x_{23}| |x_{31}|}\,,
\label{odd-basic}
\end{equation}
and the two additional ones related to it by cyclic permutations. It is clear that these are Poincar\'e invariant and homogeneous of degree zero under the dilatation $D$. Moreover, a short computation using (\ref{R-lam}), (\ref{R-bfx}) shows that they transform under the inversion $R$ into minus themselves. Therefore they are conformally invariant.\footnote{Recall that the special conformal generators may be written in terms of translations and inversion
as $K_{\mu} = RP_{\mu}R$.} It is also apparent that they are odd under the action of parity $x_k^{\mu}\rightarrow -x_k^{\mu}$, $\lambda_k\rightarrow i \lambda_k$. Therefore we can use (\ref{odd-basic}) and its cyclic permutations as conformally invariant parity odd building blocks in 3-point functions. For later convenience, we define our basis of odd tensor structures in the following form
\ie
&S_{1}=i \frac{\left(\lambda_3 {\bf x}_{31} {\bf x}_{12}\lambda_2\right)}{|x_{12}||x_{23}| |x_{31}|}
\left(\lambda_2 {\bf\check x}_{23}\lambda_3\right),\\
&S_{2}=i \frac{\left(\lambda_1 {\bf x}_{12} {\bf x}_{23}\lambda_3\right)}{|x_{12}||x_{23}| |x_{31}|}
\left(\lambda_3 {\bf\check x}_{31}\lambda_1\right),\\
&S_{3}=i \frac{\left(\lambda_2 {\bf x}_{23} {\bf x}_{31}\lambda_1\right)}{|x_{12}||x_{23}| |x_{31}|}
\left(\lambda_1 {\bf\check x}_{12}\lambda_2\right).
\label{odd-basis}
\fe
Here we have just multiplied the basic odd structures (\ref{odd-basic}) by the conformally invariant even structures $P_i$ in (\ref{even-structures}), which turns out to be slightly more convenient for our purposes since we restrict to integer spin operators in this paper. To make more manifest the fact that these tensor structures are parity odd, we can convert (\ref{odd-basis}) in vector notation by using $\lambda_{\alpha}\lambda_{\beta}=\epsilon_{\mu}\sigma^{\mu}_{\alpha\beta}$ and $\sigma_{\mu}\sigma_{\nu}=\delta_{\mu\nu}+i\epsilon_{\mu\nu\rho}\sigma_{\rho}$. This yields
\ie
&S_{1}=\frac{4\epsilon_{\mu\nu\rho}}{|x_{12}| |x_{23}|^3 |x_{31}|}  \left[x_{23}^{\mu} x_{12}^{\nu} \varepsilon_2^{\rho}\,\,\vec{\varepsilon}_3\cdot \vec{x}_{23}-\frac{1}{2} (|x_{12}|^2 x_{23}^{\mu}+|x_{23}|^2 x_{12}^{\mu}) \varepsilon_2^{\nu}\varepsilon_3^{\rho}\right],\\
&S_{2}=\frac{4\epsilon_{\mu\nu\rho}}{|x_{12}| |x_{23}| |x_{31}|^3}  \left[ x_{31}^{\mu} x_{23}^{\nu} \varepsilon_3^{\rho}\,\,\vec{\varepsilon}_1\cdot \vec{x}_{31}-\frac{1}{2} (|x_{23}|^2 x_{31}^{\mu}+|x_{31}|^2 x_{23}^{\mu}) \varepsilon_3^{\nu}\varepsilon_1^{\rho}\right],\\
&S_{3}=\frac{4\epsilon_{\mu\nu\rho}}{|x_{12}|^3 |x_{23}| |x_{31}|}  \left[x_{12}^{\mu} x_{31}^{\nu} \varepsilon_1^{\rho}\,\,\vec{\varepsilon}_2\cdot \vec{x}_{12}-\frac{1}{2} (|x_{31}|^2 x_{12}^{\mu}+|x_{12}|^2 x_{31}^{\mu}) \varepsilon_1^{\nu}\varepsilon_2^{\rho}\right],
\label{odd-vector}
\fe
which highlights the fact that they involve the antisymmetric tensor $\epsilon_{\mu\nu\rho}$. 
An alternative way to find these structures is to use the embedding formalism, see e.g. \cite{Costa:2011mg}, which leads precisely to the form (\ref{odd-vector}) in vector notations.\footnote{We thank J. Penedones for a very useful discussion.}

Note that, for instance, $S_1$ is proportional to the correlator of a scalar operator at $x_1$ and two spin-1 currents at $x_2$ and $x_3$ contracted with polarization vectors $\varepsilon_2$ and $\varepsilon_3$ in the theory of a free massless fermion
\ie
\langle \left(\bar \psi \psi\right)(x_1) \left(\bar\psi \varepsilon_2 \cdot \gamma \psi\right)(x_2)
\left(\bar\psi \varepsilon_3 \cdot \gamma \psi\right)(x_3) \rangle_{\rm free} = \frac{c}{x_{12}^2 x_{31}^2} S_1\,,
\label{110-fermi}
\fe
where $c$ is some normalization constant. In the free fermion theory the scalar operator $\bar\psi \psi$ is parity odd, so in this case this is a parity preserving 3-point function (of course, we cannot break parity in a free theory). However, more generally (\ref{110-fermi}) gives an allowed structure for the 3-point function $\langle {\cal O}_{\Delta=2}(x_1)J_{\mu}(x_2)J_{\nu}(x_3) \rangle$ in a general CFT, where ${\cal O}_{\Delta=2}$ is a dimension 2 scalar operator and $J_{\mu}$ a conserved current (see Section \ref{3pt-scalar} below for more details). If ${\cal O}_{\Delta}$ is parity even, then (\ref{110-fermi}) represents an allowed parity breaking 3-point function which can be generated in interacting CFT's which break parity.\footnote{As a test, we can check the consistency of our odd structures $S_i$ with the OPE expansion by taking certain coincidence limits. For instance, considering the 3-point function $\langle {\cal O}_{\Delta=2}(x_1)J_{\mu}(x_2)J_{\nu}(x_3) \rangle$ given by the right hand side of (\ref{110-fermi}), we can take the limit $x_1\rightarrow x_2$. In this limit, using the explicit form of $S_1$ from (\ref{odd-vector}), one finds $\langle {\cal O}_{\Delta=2}(x_1)J_{\mu}(x_2)J_{\nu}(x_3) \rangle\sim \epsilon_{\mu\rho}^{\ \ \sigma} \frac{x_{12}^{\rho}}{|x_{12}|^3}\left(\delta_{\nu\sigma}-2\frac{\left(x_{23}\right)_{\nu}\left(x_{23}\right)_{\sigma}}{x_{23}^2} \right)\frac{1}{|x_{23}|^4}$. This is consistent with the structure of the OPE, ${\cal O}_{\Delta=2}(x_1)J_{\mu}(x_2) \sim a_{\mu}^{\ \sigma}(x_{12}) J_{\sigma}(x_2)+\ldots$, and the two-point function of the spin 1 current, see eq.~(\ref{JJ-2pt}). Indeed, we can write $\langle {\cal O}_{\Delta=2}(x_1)J_{\mu}(x_2)J_{\nu}(x_3) \rangle\sim a_{\mu}^{\ \sigma}(x_{12}) \langle J_{\sigma}(x_2)J_{\nu}(x_3)\rangle$, with $a_{\mu}^{\ \sigma}(x_{12}) =\epsilon_{\mu\rho}^{\ \ \sigma} \frac{x_{12}^{\rho}}{|x_{12}|^3}$ a parity-odd OPE coefficient.}

By the counting of degrees of freedom given above, we expect that there should be some non-linear relations between the $S_i$'s and the $P_i,Q_i$'s, that are even in the $S_i$'s. This is intuitively clear, since a product of two antisymmetric tensors $\epsilon_{\mu\nu\rho}$ can be expressed in terms of the metric. Indeed, one finds the following explicit relations
\begin{eqnarray}
\label{SSrel}
&&S_3^2 = P_3^2 (Q_1 Q_2-P_3^2)\,,\qquad
S_1^2= P_1^2 (Q_2 Q_3-P_1^2)\,,\qquad
S_2^2 =  P_2^2 (Q_3 Q_1-P_2^2)\,,\cr
&&S_3S_1=P_1 P_3 (P_1 P_3-P_2 Q_2)\,,\qquad
S_2 S_3=  P_2 P_3 (P_2 P_3-P_1 Q_1)\,,\\
&&S_1 S_2=  P_1 P_2 (P_1 P_2-P_3 Q_3)\,.\nonumber
\end{eqnarray}
Therefore, while one still needs these $S_{i}$'s to build the general conformally invariant 3-point function, it is sufficient to consider structures which are at most linear in $S_{i}$. However, we observe that the counting of independent tensor structures in the 3-point functions is further complicated by the existence of the following additional identities
\begin{eqnarray}
\label{Srel}
&&(P_1^2 Q_1-P_2^2 Q_2) S_3+ (P_3^2 - Q_1 Q_2) (Q_1 S_1 - Q_2 S_2)=0\cr
&&(P_1^2 Q_1 - P_3^2 Q_3) S_2+(P_2^2 - Q_1 Q_3) (Q_1 S_1-Q_3 S_3)=0\\
&&(P_2^2 Q_2 -P_3^2 Q_3) S_1 + (P_1^2 - Q_2 Q_3) (Q_2 S_2-Q_3 S_3)=0\nonumber\,.
\end{eqnarray}
These can be seen to be true by squaring the left-hand side, using (\ref{SSrel}) and then (\ref{cubic}). Note that these are not independent conditions, since any one of them follows from the other two. Such relations, together with (\ref{cubic}), must be taken into account to write down the linearly independent structures which contribute to a given 3-point function.

To summarize, we then conclude that the most general expression for the three point function $F(\{x_i, \lambda_i, s_i \})$ consistent with conformal symmetry can be written as
\ie
F(\{x_i,\lambda_i, s_i\}) = {G_{(s_1,s_2,s_3)}(P_i, Q_i,S_{i})\over |x_{12}|^{\delta_1+\delta_2-\delta_3}|x_{23}|^{\delta_2+\delta_3-\delta_1}|x_{31}|^{\delta_3+\delta_1-\delta_2}} \,\qquad \delta_i=\Delta_i-s_i
\fe
where $G_{(s_1,s_2,s_3)}$ is a polynomial in $(P_i, Q_i,S_{i})$ which is at most linear in $S_{i}$ and is homogeneous of degree $2s_i$ in $\lambda_i$. When the currents are conserved, we should set $\delta_i=1$ in the above expression, and we must impose additional constraints coming from the conservation conditions. This will be discussed in Section 3. Below we work out some explicit examples for non-conserved currents. 


\subsection{Some simple examples}

Let us now discuss a few specific examples. For simplicity we will mostly concentrate on the case in which the currents are abelian, so we impose symmetry under pairwise exchanges when some of the spins are equal. It is straightforward to drop this assumptions and work with more general non-abelian currents.

First consider the $\langle J_1 J_1 J_1\rangle$ 3-point function. It is easy to see that if the 3-point function involve the same spin 1 operator, then this correlator is trivial by imposing symmetry under exchange of a pair of currents.\footnote{In the case of a non-abelian spin 1 current $J_1^a$, the 3-point function $\langle J_1^a J_1^b J_1^c\rangle$ is non-trivial and proportional to the structure constants $f^{abc}$ \cite{Osborn:1993cr}. The parity even conserved structures can be obtained from free theories, see the generating function (\ref{non-ab-gen}) below. Also in this case we find a new parity violating term which was not given in \cite{Osborn:1993cr}, see the Appendix.} Indeed, while there are potential conformal structures with the right degree, such as $P_1 P_2 P_3$, $Q_1 Q_2 Q_3$, $S_1 Q_1$, $S_2 Q_2$, $S_3 Q_3$, it is not possible to build out of them a structure invariant under the exchanges $(x_1,\lambda_1)\leftrightarrow (x_2,\lambda_2)$, $(x_1,\lambda_1)\leftrightarrow (x_3,\lambda_3)$, $(x_3,\lambda_3)\leftrightarrow (x_2,\lambda_2)$. For example, notice that under $(x_1,\lambda_1)\leftrightarrow (x_2,\lambda_2)$, the basic structures transform as
\begin{eqnarray}
&&P_1 \rightarrow -P_2,\quad P_2 \rightarrow -P_1,\quad P_3 \rightarrow -P_3, \quad Q_1 \rightarrow -Q_2,
\quad Q_2\rightarrow -Q_1,\quad Q_3\rightarrow -Q_3\cr
&&S_1\rightarrow S_2,\quad S_2\rightarrow S_1,\quad S_3\rightarrow S_3\,.
\end{eqnarray}

As a next example, consider $\langle J_2 J_1 J_1 \rangle$. In this case one finds 8 possible tensor structures symmetric under $(x_2,\lambda_2)\leftrightarrow (x_3,\lambda_3)$, 5 of them being parity even and 3 parity odd. Upon using the relations (\ref{cubic}) and (\ref{Srel}), these can be reduced to the following 6 linearly independent structures\footnote{To reduce the number of parity odd structures, note that (\ref{cubic}) and (\ref{Srel}) can be shown to imply the identity $Q_1 Q_2 S_2+Q_1 Q_3 S_3-2 P_2^2 S_3 -2 P_3^2 S_2 -Q_1^2 S_1 = 0$ and its cyclically related analogs.}
\ie
\langle J_2 J_1 J_1 \rangle:\quad & P_1^2 Q_1^2  \quad P_2^2 P_3^2 \quad Q_1^2 Q_2 Q_3\quad P_1 P_2 P_3 Q_1 \\
& Q_1^2 S_1\quad P_2^2 S_3+P_3^2 S_2 \,.
\label{211-conf}
\fe
Imposing that $J_2$ and $J_1$ are conserved further restricts the possible independent tensor structures. This is explained in the next section. Here we list some further examples of the structures one gets by imposing conformal invariance alone.

Consider now $\langle J_2 J_2 J_1 \rangle$. One can see from the generating functions that this correlator vanishes in the free scalar and free fermion theory.  This is, in fact, true in general if $J_2$ and $J_1$ are conserved abelian currents, see next section. Without imposing conservation, the independent conformal invariants symmetric under $(x_1,\lambda_1)\leftrightarrow (x_2,\lambda_2)$ are in this case the following 4 structures (here we used (\ref{Srel}) to eliminate one linearly dependent structure)
\ie
\langle J_2 J_2 J_1 \rangle:\quad & Q_1 Q_2 (P_1^2 Q_1-P_2^2 Q_2) \quad P_3^2(P_1^2 Q_1-P_2^2 Q_2) \\
& S_3 (P_1^2 Q_1-P_2^2 Q_2)\quad P_3^2(S_1  Q_1-S_2 Q_2)\,.
\fe

Let us analyze next the 3-point function of a spin 2 operator, $\langle J_2 J_2 J_2 \rangle$. In this case one finds 7 structures which do not involve the parity odd $S_{i}$'s, plus 5 more structures linear in $S_{i}$. Using the cubic relation (\ref{cubic}) and the identities (\ref{Srel}), these can be reduced to the following independent 4 parity even plus 2 parity odd structures
\ie
\langle J_2 J_2 J_2 \rangle:\quad & P_1^2 P_2^2 P_3^2 \quad Q_1^2 Q_2^2 Q_3^2\quad P_1 P_2 P_3 Q_1 Q_2 Q_3\quad P_1^4 Q_1^2+P_2^4 Q_2^2 +P_3^4 Q_3^2 \\
&P_1^2 Q_1^2 S_1+P_2^2 Q_2^2 S_2+P_3^2 Q_3^2 S_3\quad ~~
P_1^2 P_2^2 S_3+P_2^2 P_3^2 S_1+P_3^2 P_1^2 S_2\,.
\fe

Let us now look at some higher spin example. Consider for example the case $\langle J_4 J_1 J_1\rangle$. 
In this case, we find the following 4+2 independent conformal invariant structures (again, we need to use (\ref{cubic}) and (\ref{Srel}) to eliminate some linearly dependent structures)
\ie
\langle J_4 J_1 J_1 \rangle:\quad & P_1^2 Q_1^4 \quad P_2^2 P_3^2 Q_1^2 \quad Q_1^4 Q_2 Q_3 \quad P_1 P_2 P_3 Q_1^3 \\
& Q_1^4 S_1 \qquad P_2^2 Q_1^2 S_3+P_3^2 Q_1^2 S_2\,.
\label{}
\fe
Note that these structures are the same as the ones for $\langle J_2 J_1 J_1 \rangle$ in (\ref{211-conf}), except that they are multiplied by $Q_1^2$. It is not difficult to see that this pattern persists for $\langle J_s J_1 J_1 \rangle$, with $s$ arbitrary even spin, namely one gets the same set of structures as $\langle J_2 J_1 J_1 \rangle$ multiplied by $Q_1^{s-2}$. For $s$ odd, one can also see that the number of independent structures in $\langle J_s J_1 J_1 \rangle$ does not increase with $s$. In this case, the conformally invariant structures symmetric under the exchange $(x_2,\lambda_2) \leftrightarrow (x_3,\lambda_3)$ are
\ie
\langle J_s J_1 J_1 \rangle:\quad & Q_1^{s-1} \left(P_2^2 Q_2-P_3^2 Q_3\right) \qquad \quad s=3,5,7,\ldots\\
&Q_1^{s-2}\left(P_3^2 S_2 -P_2^2 S_3\right) \quad Q_1^{s-1} \left(Q_3 S_3 - Q_2 S_2\right)
\label{s11-odd}
\fe

As a last more complicated example we can consider $\langle J_4 J_2 J_1\rangle$. In this case we find the following 15 independent conformal invariant structures
\ie
\langle J_4 J_2 J_1 \rangle: \quad & Q_1^4 Q_2^2 Q_3 \quad Q_1^3 Q_2 Q_3 P_3^2 \quad Q_1^2 Q_3 P_3^4 \quad Q_1^3 Q_2^2 P_2^2\\
& Q_1^2 Q_2 P_2^2 P_3^2 \quad Q_1 P_2^2 P_3^4 \quad Q_1^4 Q_2 P_1^2 \quad Q_1^3 P_1^2 P_3^2\\
& Q_1^3 P_1^2 S_3\quad Q_1^4 Q_2 S_1\quad Q_1^3 P_3^2 S_1\quad Q_1^3 Q_2^2 S_2\\
& Q_1^2 Q_2 S_2\quad Q_1 P_2^2 P_3^2 S_3 \quad Q_1 P_3^4 S_2\,.
\label{421-conf}
\fe
One can proceed analogously to determine the allowed tensor structures for further examples of 3-point functions. We will not write here explicitly the list of conformally invariant tensor structures in other cases, which rapidly increases with the spins, but in the next section we will give the final results after imposing current conservation.
Some examples involving the scalar operators also will be presented in Section \ref{3pt-scalar}.

\section{Three point functions of conserved higher spin currents}

\subsection{Summary of results}
\label{summary}

Now, we want to investigate the consequence of imposing the current conservation condition $\partial^\nu J_{\nu \mu_1\cdots\mu_{s-1}}=0$. In terms of the spinorial generating function, this amounts to
\ie
(\partial_{\lambda} {\slash\!\!\!\partial_{x}} \partial_{\lambda}) J_s(x,\lambda) = 0.
\fe
If we impose the conservation law on the current $J_{s_i}$, then the correlation function obeys (up to contact terms which we neglect)
\ie
(\partial_{\lambda_i} {\slash\!\!\!\partial_{x_i}} \partial_{\lambda_i})F(\{x_j,\lambda_j,s_j\}) = 0.
\label{Ward}
\fe
After expressing $F$ in terms of $P_i, Q_i, S_i$ as in the previous section, it is unclear to us how to solve the current conservation condition in general. Nevertheless, it is straightforward to examine this case by case. This can be done for example as follows. First, we write a given 3-point function as a linear combination of conformally invariant independent structures, as explained in the previous section
\ie
F(\{x_j,\lambda_j,s_j\})=\frac{1}{|x_{12}||x_{23}||x_{31}|}\sum_k a_k {\cal T}_k(P_i,Q_i,S_i)\,,
\fe
where ${\cal T}_k$ denote a basis of linearly independent tensor structures. After evaluating (\ref{Ward}), it is convenient to use conformal invariance to set e.g. $x_1=-x_2=(1,0,0)$, $x_3=(0,0,0)$. Expanding the equation (\ref{Ward}) in powers of polarization spinors and requiring that each monomial vanishes, we get a homogeneous linear system of equations for the coefficients $a_k$. Its non-trivial solutions (if any) determine the independent conserved tensor structures for the given 3-point function. 

It is straightforward to implement this procedure using, for instance, \textit{Mathematica}. Employing this approach, we have explicitly analyzed 3-point functions of conserved currents for spins $(s_1,s_2,s_3) \le 6$, as well as some cases with two fixed low spins and one arbitrary spin. Some concrete example will be discussed in more detail below.

Based on the large amount of available data, we suggest that the three point function of {\it conserved} currents $J_s$ of spin $s$ in a three dimensional CFT have the following structure
\ie
& \langle J_{s_1} J_{s_2} J_{s_3}\rangle = a_1 \langle J_{s_1} J_{s_2} J_{s_3}\rangle_{B} + a_2 \langle J_{s_1} J_{s_2} J_{s_3}\rangle_{F} +b \langle J_{s_1} J_{s_2} J_{s_3}\rangle_{\rm odd}.
\fe
Here $\langle JJJ\rangle_B$, $\langle JJJ\rangle_F$ preserve parity, whereas $\langle JJJ\rangle_{\rm odd}$ violates parity. $a_1, a_2$ and $b$ are constant coefficients which depend on the particular set of currents appearing on the RHS. $\langle JJJ\rangle_B$ and $\langle JJJ\rangle_F$ are given in terms of the generating functions
\ie
& {\cal F}_B = \frac{1}{|x_{12}||x_{23}||x_{31}|} \exp\left[\frac{1}{2}(Q_1+Q_2+Q_3)\right]\, \cosh P_1\, \cosh P_2\, \cosh P_3\,, \\
&{\cal F}_F = \frac{1}{|x_{12}||x_{23}||x_{31}|} \exp\left[\frac{1}{2}(Q_1+Q_2+Q_3)\right]\, \sinh P_1\, \sinh P_2\, \sinh P_3\,,
\fe
expanded to order $\lambda_1^{2s_1}\lambda_2^{2s_2}\lambda_3^{2s_3}$. The structure $\langle J_{s_1} J_{s_2} J_{s_3}\rangle_{\rm odd}$, on the other hand, is conjectured to be unique when the three spins obey triangular inequality, namely $s_1\leq s_2+s_3$, $s_2\leq s_3+s_1$, $s_3\leq s_1+s_2$, and vanishes when the triangular inequality is violated. We find that this structure holds for general nonabelian higher spin currents, and does not require any assumption on exchange symmetries among currents of the same spin. In the case where there is only one current of a certain spin, for instance, some of these structures are disallowed by the symmetry that exchanges two currents in the correlator.

When $s_1+s_2+s_3=even$, $\langle JJJ\rangle_B$ and $\langle JJJ\rangle_F$ are proportional to the corresponding three point function of currents in the free theory of a massless scalar and that of a massless fermion. When $s_1+s_2+s_3=odd$, the structures  $\langle JJJ\rangle_B$ and $\langle JJJ\rangle_F$ do not show up in the free theory of a single scalar or a single fermion, but do appear in correlators of nonabelian currents in the free theories of several scalars or several fermions. For instance, suppose we take $NM$ free complex massless scalars or fermions, and consider $U(N)$ invariant higher spin currents $J_s^a$, which are bilinears in the fields and take value in the global $U(M)$ flavor symmetry algebra. Their three point functions are given in terms of the generating functions\footnote{This can be derived analogously to the appendix of \cite{Giombi:2010vg}.}
\ie
& {\cal F}_{boson}^{abc} =  \frac{d^{abc}\cosh\left[\frac{1}{2}(Q_1+Q_2+Q_3)\right]
+ f^{abc}\sinh\left[\frac{1}{2}(Q_1+Q_2+Q_3)\right]}{|x_{12}||x_{23}||x_{31}|}  \cosh P_1\, \cosh P_2\, \cosh P_3\,, \\
& {\cal F}_{fermion}^{abc} =  \frac{d^{abc}\sinh\left[\frac{1}{2}(Q_1+Q_2+Q_3)\right]
+ f^{abc}\cosh\left[\frac{1}{2}(Q_1+Q_2+Q_3)\right]}{|x_{12}||x_{23}||x_{31}|}  \sinh P_1\, \sinh P_2\, \sinh P_3\,. \\
\label{non-ab-gen}
\fe
Here $d^{abc}={\rm Tr}(T^a T^{(b} T^{c)})$, $f^{abc} = {\rm Tr}(T^a T^{[b} T^{c]})$, where $T^a$ are the generators of $U(M)$ flavor symmetry.

While we do not know a closed form formula for $\langle J_{s_1} J_{s_2} J_{s_3}\rangle_{\rm odd}$ with general spins, in all the cases we have examined, we find that such a structure exists and is unique when the triangular inequalities on the spins are obeyed, and does not exist when triangular inequalities are violated. We conjecture that this statement is true for all spins.\footnote{After the original version of this paper appeared, a generating function for the parity odd 3-point functions, which incorporates the triangle rule, was given in \cite{MZ1}. However, a general proof that these are the only allowed structures is still lacking.}  Note that the triangular selection rule only applies to the case where the 3-point function obeys conservation on all three currents. If $s_1>s_2+s_3$, one can have conformally invariant parity odd tensor structures which are conserved in $J_{s_2}$ and $J_{s_3}$, but not $J_{s_1}$ (see for instance eq. (\ref{JsJ1J1}) and discussion thereafter). In the next subsection we will present some explicit formulae for these odd structures in some special cases. 

The $\langle JJJ\rangle_{odd}$ structures do not show up in free field theories.  They can arise in interacting three dimensional CFTs that violate parity. For example they are generically non-zero at odd loop order in the theory of a fundamental fermion or boson coupled to a $U(N)$ Chern-Simons gauge field, as demonstrated in \cite{GMPSY, Ofer-et-al}. 

\subsection{Examples}

Let us start from the $\langle J_2 J_1 J_1 \rangle$ example discussed above. We write the correlator as a linear combination of the 6 conformally invariant structures given in (\ref{211-conf})
\ie
\langle J_2 J_1 J_1 \rangle=\frac{1}{|x_{12}||x_{23}||x_{31}|}&\left[a_1 P_1^2 Q_1^2+a_2 P_2^2 P_3^2+a_3 Q_1^2Q_2Q_3+a_4 P_1 P_2 P_3 Q_1\right. \\
& \left.b_1 Q_1^2 S_1+b_2 (P_2^2 S_3+P_3^2 S_2)\right]
\label{211-conserved}
\fe
Imposing conservation one finds the conditions $a_2=-4a_1$, $a_3=-5/2 a_1$ and $b_2=2b_1$, so the end result is that conformal symmetry and current conservation restrict the 3-point function to the form
\ie
\langle J_2 J_1 J_1 \rangle=\frac{1}{|x_{12}||x_{23}||x_{31}|}&\left[a_1 \left(P_1^2 Q_1^2-4P_2^2 P_3^2-\frac{5}{2} Q_1^2Q_2Q_3\right)+a_4 P_1 P_2 P_3 Q_1\right. \\
& \left.+b_1 \left(Q_1^2 S_1+2 P_2^2 S_3+2P_3^2 S_2\right)\right]\,.
\fe
As in \cite{Osborn:1993cr}, the two parity even structures in the first line are linear combinations of free scalar and free fermion structures, see eq. (\ref{free-gen}) (the term proportional to $a_4$ corresponds to free fermions, while the term proportional to $a_1$ is a linear combination of free scalars and free fermions). On the other hand, the parity odd structure in the second line was not given in the analysis by Osborn and Petkou \cite{Osborn:1993cr}.

In the case of $\langle J_2 J_2 J_1 \rangle$, we start from a linear combination of the 4 structures found in the previous section
\ie
\langle J_2 J_2 J_1 \rangle =\frac{1}{|x_{12}||x_{23}||x_{31}|}&\left[a_1\, Q_1 Q_2 (P_1^2 Q_1-P_2^2 Q_2)  + a_2\, P_3^2(P_1^2 Q_1-P_2^2 Q_2)\right.  \\
&\left. +b_1\,S_{3} (P_1^2 Q_1-P_2^2 Q_2)+ b_2\,P_3^2(S_{1}  Q_1-S_{2} Q_2)\right]\,,
\fe
and imposing conservation we find that $a_1=a_2=b_1=b_2=0$. Therefore we conclude that the correlator $\langle J_2 J_2 J_1 \rangle$ vanishes in any CFT in which the currents $J_2$ and $J_1$ are conserved. Note that here we have assumed the two $J_2$'s to be the same spin-2 current, and have imposed symmetry under the exchange of the two currents on the RHS. In theories with more than one conserved spin-2 currents, the correlator of two spin-2 and one spin-1 currents can be nontrivial, as discussed in the previous subsection.

Next, let us look at the 3-point function of the stress tensor $T=J_2$. In this case, following the analysis above, we write
\ie
\langle J_2 J_2 J_2 \rangle= \frac{1}{|x_{12}||x_{23}||x_{31}|}& \left[ a_1\, P_1^2 P_2^2 P_3^2 + a_2\, Q_1^2 Q_2^2 Q_3^2 + a_3\,P_1 P_2 P_3 Q_1 Q_2 Q_3 \right. \\
& + a_4\, (P_1^4 Q_1^2+P_2^4 Q_2^2 +P_3^4 Q_3^2) \\
&+ b_1 (P_3^2 Q_3^2 S_{3}+P_1^2 Q_1^2 S_{1}+P_2^2 Q_2^2 S_{2}) \\
&\left.+ b_2 (P_1^2 P_2^2 S_{3}+P_2^2 P_3^2 S_{1}+P_3^2 P_1^2 S_{2})\right]\,.
\fe
The conservation equations now yield the conditions
\ie
a_3=\frac{5}{4}a_1-\frac{16}{5}a_2\,,\qquad a_4=-\frac{8}{15}a_2\,,\qquad
b_2 = 5 b_1\,.
\fe
Therefore we find that there are 2 independent parity even structures, plus one additional parity odd structure. The two parity preserving structures can be seen to be a linear combination of free scalars and free fermions results, which agrees with \cite{Osborn:1993cr}. The additional parity odd conserved structure
\ie
\langle J_2 J_2 J_2 \rangle_{\rm parity~odd}=\frac{(P_1^2 Q_1^2 +5 P_2^2 P_3^2)S_{1}+(P_2^2 Q_2^2+5P_3^2 P_1^2) S_{2}+(P_3^2 Q_3^2+5P_1^2 P_2^2) S_{3}}{|x_{12}||x_{23}||x_{31}|}
\fe
is new, and it can appear for instance in parity violating Chern-Simons matter theories \cite{GMPSY}.

Next consider the $\langle J_s J_1 J_1 \rangle$ correlator. We should distinguish the cases with  $s$ even and $s$ odd. In the former case, we start from (see (\ref{211-conf}) and the following comments)
\ie
\langle J_s J_1 J_1 \rangle= \frac{1}{|x_{12}||x_{23}||x_{31}|}& Q_1^{s-2}\left[a_1\, P_1^2 Q_1^2+ a_3\, P_2^2 P_3^2+ a_2\, Q_1^2 Q_2 Q_3+ a_4\, P_1 P_2 P_3 Q_1 \right.\\
&\left.+ b_1\, Q_1^2 S_{1}+ b_2\, (P_2^2 S_{3}+P_3^2 S_{2})\right]\,. \qquad s=2,4,6,\ldots
\fe
Imposing conservation on the spin 1 current, we get the constraints
\ie
&a_2 =-2 s a_1 \,,\quad a_3=-\frac{2s+1}{2s-2} a_1\\
&b_2=s b_1
\fe
hence
\ie
\langle J_s J_1 J_1 \rangle= \frac{1}{|x_{12}||x_{23}||x_{31}|}&Q_1^{s-2}\Big{[}a_1\, (P_1^2 Q_1^2-\frac{2s+1}{2s-2}\, Q_1^2 Q_2 Q_3 -2 s \, P_2^2 P_3^2)+ a_4\, P_1 P_2 P_3 Q_1 \\
&+ b_1\, (Q_1^2 S_{1}+ s\,P_2^2  S_{3}+s\,P_3^2  S_{2})\Big{]}\qquad s=2,4,6,\ldots
\label{JsJ1J1}
\fe
The two independent structures that do not involve $S_{i}$ are linear combinations of free scalars and free fermions results (the structure proportional to $a_4$ is precisely the free fermion answer, while the one proportional to $a_1$ is a linear combination of fermions and bosons results, as can be checked from (\ref{free-gen})). Notice that even though we have not yet imposed conservation on $J_s$, this is automatically satisfied by these two structures. On the other hand, the structure which contains $S_{i}$ in the second line above does not satisfy conservation on $J_s$ for $s\ge 4$ unless $b_1$=0, therefore we conclude that it cannot appear in a theory in which $J_s$, $s\ge 4$ is exactly conserved.\footnote{However, it is expected that this structure can appear for example at odd loop order in the theory of a fundamental fermion coupled to CS gauge field. For $s=4$, this structure can be related to an anomalous Ward identity obeyed by the 3-point function $\langle J_4 J_1 J_1\rangle$, which follows from the fact that the divergence of $J_4$ mixes with ``multi-trace" operators, including in particular $J_1 J_1$ \cite{GMPSY}.} The case of $\langle J_s J_1 J_1 \rangle$ with odd $s$ can be studied similarly, starting from (\ref{s11-odd}). Imposing conservation on $J_1$, one finds that the parity even structure in (\ref{s11-odd}) is not permitted, and the relative coefficients of the two parity odd structures is fixed as
\ie
\langle J_s J_1 J_1 \rangle = \frac{b}{|x_{12}||x_{23}||x_{31}|}&Q_1^{s-2} \Big{[}
(Q_1 Q_2+s P_3^2)S_2 -(Q_1 Q_3+s P_2^2) S_3\Big{]}\qquad s=3,5,\ldots
\fe
Note that since this is a parity breaking structure, it does not arise in particular in free theories. If one further imposes conservation on $J_s$, one finds that it is only possible if $b=0$, hence this 3-point function vanishes in a 3d CFT with conserved $J_s$ ($s$ odd).


For $\langle J_4 J_2 J_1\rangle$, starting from a linear combination of the 15 conformally invariant structures listed in (\ref{421-conf}) and imposing conservation on all currents, we end up with the following independent conserved structures\footnote{Here we have used (\ref{cubic}) to rewrite the tensor structures in a simpler form.}
\ie
\langle J_4 J_2 J_1 \rangle =& \frac{1}{|x_{12}||x_{23}||x_{31}|}\Big{[}a_1 \left( Q_1^4 Q_2^2 Q_3-\frac{64}{15}  Q_1 P_2^2 P_3^4-\frac{64}{15} Q_1^3 P_1^2 P_3^2  -\frac{128}{15} Q_1^2 Q_2 P_2^2 P_3^2\right. \\
&\left. -\frac{4}{3} Q_1^4 Q_2 P_1^2
-\frac{8}{5} Q_1^3 Q_2^2 P_2^2 -\frac{16}{5} Q_1^2 Q_3 P_3^4  \right)
+ a_2 P_1 P_2 P_3 Q_1^2 (Q_1 Q_2+2 P_3^2)\Big{]}\,.
\label{421-cons}
\fe
Note that in this case all the parity odd structures in (\ref{421-conf}) are killed by imposing conservation on the currents.\footnote{Two independent linear combinations of the parity odd structures in (\ref{421-conf}) would be allowed if one imposes conservation only on $J_2$ and $J_1$.} This is in agreement with the claim in the previous subsection that the parity odd structure does not exist when the triangular inequalities on the spins are violated.
Observe that the two conserved structures found above can be obtained by expanding to the appropriate powers of polarization spinors the two following objects
\ie
&\frac{1}{|x_{12}||x_{23}||x_{31}|} \sinh\left[\frac{1}{2}(Q_1+Q_2+Q_3)\right]\, \cosh P_1\, \cosh P_2\, \cosh P_3 \\
&\frac{1}{|x_{12}||x_{23}||x_{31}|} \cosh\left[\frac{1}{2}(Q_1+Q_2+Q_3)\right]\, \sinh P_1\, \sinh P_2\, \sinh P_3\,,
\label{pseudo-gen}
\fe
which are just the terms proportional to $f^{abc}$ in the generating functions of 3-point correlators of non-abelian currents in free theories, see eq. (\ref{non-ab-gen}).
In particular expanding the second line to order $\lambda_1^8 \lambda_2^4 \lambda_3^2$ one gets the structure proportional to $a_2$ in (\ref{421-cons}), while expanding the first line gives a linear combination of the $a_1$ and $a_2$ structures in (\ref{421-cons}). It is easy to verify that the expansion of (\ref{pseudo-gen}) to the power $\lambda_1^{2 s_3} \lambda_2^{2s_2} \lambda_3^{2s_3}$ generates new non-trivial conformal invariant and conserved structures in the case $\sum s_i = {\rm odd}$.  Note however that (\ref{pseudo-gen}) are antisymmetric under any pairwise interchange $(x_i,\lambda_i) \leftrightarrow (x_j,\lambda_j)$, hence they cannot be used to generate 3-point structures for abelian currents when any two of the spins are equal.
So our conclusion, as summarized in Section \ref{summary}, is that for $\langle J_{s_1} J_{s_2} J_{s_3}\rangle$ with $\sum s_i = {\rm odd}$ and $s_1\neq s_2\neq s_3$ there are two parity even conformally invariant and conserved structures besides those of a free massless scalar/fermion. By looking at several further examples, it appears that (\ref{free-gen}) and (\ref{pseudo-gen}) exhaust all possible parity even conserved structures.

We end this section by listing the results for the correlators $\langle J_4 J_2 J_2\rangle$, $\langle J_4 J_4 J_2\rangle$ and $\langle J_4 J_4 J_4 \rangle$, obtained after imposing conformal invariance together with conservation on all currents. We find
\begin{eqnarray}
&&\langle J_4 J_2 J_2 \rangle = a_1 \langle J_4 J_2 J_2 \rangle_{\rm free~scalars} + a_2 \langle J_4 J_2 J_2 \rangle_{\rm free~fermions} \cr
&&+ b_1\frac{Q_1}{|x_{12}||x_{23}||x_{31}|}\left[2 S_1 Q_1 \left(P_1 P_2 P_3 Q_1 + P_1^2 Q_1^2-6P_2^2 P_3^2\right)
\right.\\
&&\left.~~~~~~~~~~~~~~~~~~~~~~~
-(S_2 Q_2+S_3 Q_3) \left( 6 P_2^2 P_3^2 + 5 Q_1^2 Q_2 Q_3\right)
\right]\,.\cr
\cr
&&\langle J_4 J_4 J_2 \rangle = a_1 \langle J_4 J_4 J_2 \rangle_{\rm free~scalars} + a_2 \langle J_4 J_4 J_2 \rangle_{\rm free~fermions} \cr
&&+b_1 \frac{1}{|x_{12}||x_{23}||x_{31}|}\Big{[} Q_1^4 Q_2^3 Q_3 S_1 +Q_1^3 Q_2^4 Q_3 S_2
+\frac{1}{10} \Big{(}126 P_1^3 P_2 P_3 Q_1^2 Q_2\cr
&&-14 P_1^4 Q_1^2 (P_3^2 + 3 Q_1 Q_2)+P_1^2 Q_1 Q_3(24 P_3^4 - 37 P_3^2 Q_1 Q_2 - 117 Q_1^2 Q_2^2) \\
&&-\frac{1}{2}Q_3^2(2 P_3^6 + 67 P_3^4 Q_1 Q_2 - 10 P_3^2 Q_1^2 Q_2^2 - 189 Q_1^3 Q_2^3)+(x_1,\lambda_1)\leftrightarrow(x_2,\lambda_2)\Big{)}
S_3\Big{]}\,.\cr
\cr
&&\langle J_4 J_4 J_4 \rangle = a_1 \langle J_4 J_4 J_4 \rangle_{\rm free~scalars} + a_2 \langle J_4 J_4 J_4 \rangle_{\rm free~fermions} \cr
&&+b_1 \frac{1}{|x_{12}||x_{23}||x_{31}|}\Big{[}Q_1 S_1 \Big{(}
16 P_1^6 Q_1^3 + 128 P_2^6 Q_2^3 + 128 P_3^6 Q_3^3 + 8 P_1^4 Q_1^3 Q_2 Q_3 \\
&&- 608 P_2^4 Q_1 Q_2^3 Q_3-
 608 P_3^4 Q_1 Q_2 Q_3^3 -414 P_1^2 Q_1^3 Q_2^2 Q_3^2 + 627 Q_1^3 Q_2^3 Q_3^3\Big{)}+{\rm cyclic}\Big{]}\,.\nonumber
\end{eqnarray}

\subsection{Examples involving a scalar operator}
\label{3pt-scalar}
The explicit examples discussed so far involve currents of spins $s_i\ge 1$. Of course, the formalism we set up above can be easily applied to the case in which the 3-point functions involve scalar primary operators. Let us imagine that the CFT under study has a scalar primary ${\cal O}_{\Delta}$ of dimension $\Delta$. In the specific example of the duality between Vasiliev's theories \cite{Vasiliev:1999ba} and vector models \cite{Sezgin:2002rt}\cite{Klebanov:2002ja}, the relevant case is $\Delta=1$ or $\Delta=2$, and the scalar operator can be either even or odd under parity.

The simplest case to consider is the 3-point function involving a higher spin current and two scalars, $\langle J_s(x_1,\lambda_1) {\cal O}_{\Delta}(x_2) {\cal O}_{\Delta}(x_3)\rangle$. In this case it is easy to see that among the conformally invariant tensor structures $(P_i,Q_i,S_i)$ we can only use $Q_1$. Therefore we immediately conclude that conformal invariance implies
\ie
\langle J_s(x_1,\lambda_1) {\cal O}_{\Delta}(x_2) {\cal O}_{\Delta}(x_3) \rangle = \frac{a}{|x_{12}||x_{23}|^{2\Delta-1}|x_{31}|}~ \left(Q_1\right)^s\,,\qquad s=2,4,6,\ldots
\fe
while it vanishes if $s$ is odd (this is because $Q_1$ changes sign if we exchange $x_2$ and $x_3$). Here $a$ is some arbitrary constant. One can see that this obeys the Ward identity for current conservation on $J_s$ for any choice of $\Delta$.

In the case of the correlator of two higher spin currents at $x_1$, $x_2$ and one scalar at $x_3$, we see that the possible building blocks are $Q_1,Q_2,P_3,S_3$. Then we have
\ie
\langle J_{s_1}(x_1,\lambda_1) J_{s_2}(x_2,\lambda_2) {\cal O}_{\Delta}(x_3)\rangle =
\frac{1}{|x_{12}|^{2-\Delta}|x_{23}|^{\Delta}|x_{31}|^{\Delta}}&\left[G(Q_1,Q_2,P_3)
+S_3 \tilde G(Q_1,Q_2,P_3) \right]\,,
\fe
where $G$ and $\tilde G$ are homogeneous of degree respectively $(2s_1,2s_2)$ and $(2s_1-2,2s_2-2)$ in $(\lambda_1,\lambda_2)$. In a parity invariant CFT, either $G$ or $\tilde G$ must vanish. If ${\cal O}_{\Delta}$ is parity even, then $\tilde G$ should vanish and conversely $G$ vanishes if ${\cal O}_{\Delta}$ is parity odd. More generally, both $G$ and $\tilde G$ can appear if parity is broken.  For example, in the free massless scalar theory with $\Delta=1$ parity even scalar, $\tilde G$=0 and the generating function for $\langle J_{s_1}J_{s_2}J_0\rangle$ is given by (\ref{free-gen}) upon setting $\lambda_3=0$. In the free fermion theory, with $\Delta=2$ parity odd scalar operator, it may be shown that a generating function for $\langle J_{s_1}J_{s_2}J_0\rangle$ of all spins can be written as\footnote{We obtained this simple generating function from Vasiliev's higher spin gauge theory in $AdS_4$. Its derivation will be presented elsewhere.}
\ie
{\cal F}^{\rm fermi}_{\langle J_{s_1}J_{s_2}J_0 \rangle} = \frac{1}{|x_{23}|^2|x_{31}|^2} ~S_3 \cosh\Big{(}\frac{Q_1+Q_2}{2}\Big{)}\, \frac{\sinh P_3}{P_3}\,.
\label{JJO-fermi}
\fe
In particular for $\langle J_1 J_1 J_0\rangle$ this reproduces the result quoted earlier.

It is straightforward to work out some specific examples. For instance, before imposing conservation, we have
\ie
\langle J_1 (x_1,\lambda_1) J_1(x_2,\lambda_2) {\cal O}_{\Delta}(x_3)\rangle
= \frac{1}{|x_{12}|^{2-\Delta}|x_{23}|^{\Delta}|x_{31}|^{\Delta}} \left[
a_1 Q_1 Q_2 +a_2 P_3^2+b S_3 \right]\,.
\fe
Requiring that $J_1$ is a conserved current yields the condition $(4-2\Delta)a_1-\Delta a_2=0$ and $b$ arbitrary, therefore we find
\ie
\langle J_1 J_1 {\cal O}_{\Delta}\rangle= \frac{1}{|x_{12}|^{2-\Delta}|x_{23}|^{\Delta}|x_{31}|^{\Delta}} \left[a \left(\Delta Q_1 Q_2+(4-2\Delta)P_3^2\right)+b S_3\right]\,.
\label{JJOdel}
\fe
For $\Delta=1$ and $b=0$ we recognize the structure of the free scalar result in eq. (\ref{free-gen}), while for $\Delta=2$ and $a=0$ we find the structure of the free fermion result, see eq. (\ref{110-fermi}). In the case of the critical scalar $U(N)$ theory (which is a parity preserving CFT), where $\Delta=2+{\cal O}(1/N)$ and ${\cal O}_{\Delta}$ is parity even, we should find $\langle J_1 J_1 {\cal O}_{\Delta=2}\rangle=a Q_1 Q_2/(x_{23}^2 x_{31}^2)$ at large $N$. In parity violating Chern-Simons matter theories, we expect that the parity preserving term can be non-zero at even loop order and the parity breaking one at odd loop order (which of the two terms in (\ref{JJOdel}) is parity breaking or preserving depends on whether the scalar operator ${\cal O}_{\Delta}$ is parity even or odd).

Going up in spin, consider $\langle J_2 J_1 {\cal O}_{\Delta}\rangle$. In this case we find that conservation on $J_2$ and $J_1$ implies that this correlator must vanish, except in the special cases $\Delta=1$ and $\Delta=2$, where we find the structures
\begin{equation}
\begin{aligned}
&\langle J_2 J_1 {\cal O}_{\Delta=1}\rangle = \frac{a}{|x_{12}||x_{23}||x_{31}|}\left( Q_1^2 Q_2 + 4 Q_1 P_3^2\right)\\
&\langle J_2 J_1 {\cal O}_{\Delta=2}\rangle = \frac{b}{x_{23}^2 x_{31}^2} Q_1 S_3\,.
\end{aligned}
\end{equation}
which are respectively parity even and parity odd. 
In the free theories of a single scalar or fermion, $\langle J_2 J_1 {\cal O}_{\Delta}\rangle$ (with $\Delta=1,2$ respectively) vanishes, as can be checked by direct computation or from the generating functions given earlier. However, note that the structure we found for $\langle J_2 J_1 {\cal O}_{\Delta=1}\rangle$ is contained in the term proportional to $f^{abc}$ of the boson generating function in eq.~(\ref{non-ab-gen}). The $\Delta=2$ parity odd structure also corresponds to the three point function of non-abelian currents in the theory of several free fermions.
 

The next non-trivial example is $\langle J_2 J_2 {\cal O}_{\Delta}\rangle$, where we find two independent conserved structures
\ie
&\langle J_2 J_2 {\cal O}_{\Delta}\rangle=\frac{1}{|x_{12}|^{2-\Delta}|x_{23}|^{\Delta}|x_{31}|^{\Delta}} \Big{[}
a \Big{(} \Delta (\Delta+2) Q_1^2 Q_2^2+\Delta(32-8\Delta) Q_1 Q_2 P_3^2 \\
&~~~~~~~~~~~~~~~
+8(6-6\Delta+\Delta^2)P_3^4\Big{)}+b~S_3\left((\Delta+1)Q_1 Q_2+(6-2\Delta)P_3^2\right)\Big{]}\,.
\fe
One may check that for $\Delta=1$, $b=0$ and $\Delta=2$, $a=0$ we recover respectively the free scalar and free fermion results, see eq. (\ref{free-gen}) and (\ref{JJO-fermi}).

Let us finally list some examples involving higher spins, namely $\langle J_3 J_1 {\cal O}_{\Delta} \rangle$, $\langle J_5 J_1 {\cal O}_{\Delta}\rangle$ and $\langle J_4 J_2 {\cal O}_{\Delta}\rangle$. Imposing conservation only on $J_1$, we find that
\ie
&\langle J_3 J_1 {\cal O}_{\Delta}\rangle=\frac{1}{|x_{12}|^{2-\Delta}|x_{23}|^{\Delta}|x_{31}|^{\Delta}}\left[
a_3 \left(\Delta Q_1^3 Q_2+(8-2\Delta)Q_1^2 P_3^2\right)+b_3 Q_1^2 S_3\right]\,,\\
&\langle J_5 J_1 {\cal O}_{\Delta}\rangle=\frac{1}{|x_{12}|^{2-\Delta}|x_{23}|^{\Delta}|x_{31}|^{\Delta}}\left[
a_5 \left(\Delta Q_1^5 Q_2+(12-2\Delta)Q_1^4 P_3^2\right)+b_5 Q_1^4 S_3\right]\,.
\fe
Exact conservation of $J_3$ then implies
\ie
\langle \partial \cdot J_3 ~J_1 {\cal O}_{\Delta}\rangle=0 ~~\leftrightarrow ~~&b_3=0,~\Delta=1\,\\
& a_3=0,~\Delta=2\,,\\
&b_3=a_3=0,~{\rm if}~\Delta\neq 1,2
\fe
and imposing conservation on $J_5$ for $\langle J_5 J_1 {\cal O}_{\Delta}\rangle$ one finds exactly the same conditions on $a_5$, $b_5$. The structures which are conserved on all currents just correspond to free scalars and free fermions. Analogous results can be written down for $\langle J_s J_1 {\cal O}_{\Delta}\rangle$, $s=7,9,\ldots$, with the total number of conserved structures not increasing with $s$. 

Finally, for $\langle J_4 J_2 {\cal O}_{\Delta}\rangle$ (similar results apply to $\langle J_s J_2 {\cal O}_{\Delta}\rangle$ for $s\ge 4$), conservation of the stress-tensor $J_2$ implies
\ie
\langle J_4 J_2 {\cal O}_{\Delta}\rangle =
\frac{1}{|x_{12}|^{2-\Delta}|x_{23}|^{\Delta}|x_{31}|^{\Delta}}&\left[
a \left(\Delta(\Delta+2) Q_1^4 Q_2^2 + 8(15-10\Delta+\Delta^2)Q_1^2 P_3^4 \right.\right.\\
&+ \left.8\Delta(7-\Delta) Q_1^3 Q_2 P_3^2\right)\\
&+\left.b ~S_3Q_1^2\left((\Delta+1)Q_1 Q_2+(10-2\Delta)P_3^2\right)\right]\,.
\fe
Analogously as above, conservation on $J_4$ requires
\ie
\langle \partial \cdot J_4 ~J_2 {\cal O}_{\Delta}\rangle=0 ~~\leftrightarrow ~~&b=0,~\Delta=1\,\\
& a=0,~\Delta=2\,,\\
&b=a=0,~{\rm if}~\Delta\neq 1,2\,.
\fe

It is interesting to observe how imposing conservation of the higher-spin currents singles out the cases $\Delta=1$, $\Delta=2$ as special, corresponding to the dimensions of scalar operators in free theories. From the AdS/CFT point of view, these are the conformal dimensions associated to a scalar field in $AdS_4$ with $m^2=-2/R_{AdS}^2$. This is indeed as expected from the structure of the 4d higher spin gauge theory \cite{Vasiliev:1999ba}.

\section{Comments on the holographic dual}

In this paper we have found new conformally invariant parity odd tensor structures for three point functions of conserved currents of various spins, and one may ask if they arise in actual three-dimensional CFTs. It will be shown in \cite{GMPSY} via explicit perturbative computations that this indeed occurs in parity violating conformal Chern-Simons-matter theories. For now, we will make a few comments on the holographic dual of such CFTs. A large $N$ CFT in three dimensions is expected to be dual to a theory of gravity in $AdS_4$. If there are in additional conserved higher spin currents in the CFT, the dual bulk theory would contain higher spin gauge fields. How parity even three point functions of higher spin currents arise from a bulk higher spin gauge theory was understood in \cite{Giombi:2009wh, Giombi:2010vg}.
Here we will consider a few possible terms in the bulk Lagrangian that could give rise to the parity odd three point functions.

Consider first the parity odd structure of the three point function of the stress energy tensor, $\langle TTT\rangle$. In the bulk Lagrangian, the parity odd term with the least number of derivatives, that is not a total derivative, and does not vanish by the equation of motion, is expressed in terms of Weyl tensor as
\ie
(W_+)^3 - (W_-)^3 = W_{\A\B}{}^{\C\D}W_{\C\D}{}^{\sigma\tau} W_{\sigma\tau}{}^{\A\B}
-W_{\da\db}{}^{\dc\dd}W_{\dc\dd}{}^{\dot\sigma\dot\tau} W_{\dot\sigma\dot\tau}{}^{\da\db}
\fe
where $\A,\B,\cdots$ are chiral spinor indices and $\da,\db,\cdots$ are anti-chiral spinor indices. This term is expected to contribute to the parity odd $\langle TTT\rangle$ structure in the boundary theory. Similarly, there is a parity violating graviton-vector field coupling
\ie
W_+ F_+^2 - W_- F_-^2 = W_{\A\B\C\D} F^{\A\B} F^{\C\D} - W_{\da\db\dc\dd} F^{\da\db} F^{\dc\dd}
\fe
that could give rise to the parity odd $\langle Tjj\rangle$ structure we found.

For higher spin gauge fields, one may consider bulk cubic terms involving fields of spins $s_1$, $s_2$, and $s_3$, of the form
\ie\label{wwwa}
\left.W_+^{(s_1)}(y)*W_+^{(s_2)}(y)*W_+^{(s_3)}(y)\right|_{y=0} - \left.W_-^{(s_1)}(\bar y)*W_-^{(s_2)}(\bar y)*W_-^{(s_3)}(\bar y)\right|_{\bar y=0},
\fe
where
\ie
W_+(y) = \sum W_{\A_1\cdots \A_{2s}} y^{\A_1}\cdots y^{\A_{2s}},~~~~
W_-(\bar y) = \sum W_{\da_1\cdots \da_{2s}} \bar y^{\da_1}\cdots \bar y^{\da_{2s}}.
\fe
Here $W_{\A_1\cdots \A_{2s}}$ are the self dual components of the generalized Weyl tensor of the spin $s$ field, and similarly $W_{\da_1\cdots \da_{2s}}$ the anti self dual components. The $*$ product is defined as in \cite{Vasiliev:1999ba}. Here it simply contracts $y^\A$ with $y^\B$, giving $\epsilon^{\A\B}$, and similarly on the $\bar y$'s. While we expect (\ref{wwwa}) to give a parity odd $\langle JJJ\rangle$ structure when the three spins $s_1,s_2,s_3$ obey triangular inequality, (\ref{wwwa}) vanishes when the three spins do not obey triangular inequality. This is precisely in agreement with the absence of parity odd, conformally invariant and conserved $\langle J_{s_1} J_{s_2} J_{s_3}\rangle$ structure, as discussed in earlier sections.\footnote{Though, naively, it appears that one can write down parity odd cubic couplings of higher spin Weyl tensors violating triangular inequality, by acting with covariant derivatives. Nonetheless, our previous analysis indicates that such terms cannot contribute to the three point function as forbidden by conformal symmetry and higher spin gauge symmetry (current conservation in the boundary theory).}
A more detailed exploration of such parity violating bulk couplings of higher spin fields and in particular their connection to Vasiliev's higher spin gauge theory in $AdS_4$ will be given elsewhere.

\subsection*{Acknowledgments}

We are grateful to S. Minwalla and S. Trivedi for collaboration on closely related topics, and to S. Hartnoll, D. Hofman and J. Penedones for discussions. We thank A. Zhiboedov for pointing out some errors in the earlier version of this paper. S.G. would like to thank the Simons Center for Geometry and Physics for hospitality during the Higher Spin Theories and Holography Workshop. S.P. would like to thank the organizers of the Indian Strings Meeting 2011 and the Indian Institute of Science, Bangalore, for hospitality. X.Y. would like to thank the organizers of Indian Strings Meeting 2011, Tata Institute of Fundamental Research, Berkeley Center for Theoretical Physics, and the Simons Center for Geometry and Physics, where some of this work was presented. S.G. is supported by Perimeter Institute
for Theoretical Physics. Research at Perimeter Institute is supported by the
Government of Canada through Industry Canada and by the Province of Ontario through
the Ministry of Research $\&$ Innovation. S.P. gratefully acknowledges the people of India for their generous support for research in the basic sciences. X.Y. is supported in part by the Fundamental Laws Initiative Fund at Harvard University, and by NSF Award PHY-0847457.

\appendix

\section{Calculation of $\langle j^a j^b j^c\rangle$ in vector notation}
\newcommand{\cor}[1]{\left\langle #1 \right \rangle }

In this appendix, we obtain the allowed terms for the correlation function $\cor{j^a_\mu(x_1) j^b_\nu(x_2) j^c_\omega(x_3)}$, which appears to be the simplest correlation function admitting a new parity-violating structure, using the formalism of Osborn and Petkou \cite{Osborn:1993cr}. This illustrates how the existence of parity-violating terms can be seen in their formalism, and  provides a simple cross-check of our approach.

Following \cite{Osborn:1993cr}, the correlation function must take the form:
\begin{equation}
 \cor{j^a_\mu(x_1) j^b_\nu(x_2) j^c_\omega(x_3)} = \frac{f^{abc}}{x_{12}^2 x_{23}^2 x_{13}^2} I_{\mu \alpha}(x_{13}) I_{\nu \beta}(x_{23}) t_{\alpha \beta \omega} (X_{12})
\end{equation}
where $X_{12} = \frac{x_{13}}{x_{13}^2} - \frac{x_{23}}{x_{23}^2}$ and $I_{\mu \alpha}(X) = \delta_{\mu \alpha} - 2 X_\mu X_\alpha X^{-2}$. These are related to the parity-invariant objects defined in the main text via:
\begin{eqnarray}
 Q_1  & = & 2 \varepsilon_1^\mu X_{23\mu} , \\
 (P_3)^2 & = & -2 \varepsilon_1^\mu \varepsilon_2^\nu x_{12}^{-2} I_{\mu \nu}(x_{12})
\end{eqnarray}

The quantity $t_{\alpha \beta \omega}$ must have scaling dimension zero and satisfy:
\begin{eqnarray}
 t_{\mu \nu \omega}(X) & = &- t_{\nu \mu \omega}(-X) \label{A4} \\
 I_{\mu \alpha} t_{\alpha \nu \omega}(X) & = & -t_{\omega \mu \nu}(X) \label{A5} \\
\left( \partial_\mu-2\frac{X_\mu}{X^2} \right) t_{\mu \nu \omega}(X) & = & 0
\end{eqnarray}
The first and second conditions ensure that the correlation function has the required symmetries, the second condition also effectively ensures that the correlation function is invariant under inversions (up to a minus sign for parity-violating terms), and the third condition is due to current conservation. The signs in equations (\ref{A4}) and (\ref{A5}), which originate from equations (2.20) and (2.21) of \cite{Osborn:1993cr}, have been carefully arranged to allow both parity-preserving and parity-violating solutions.

The general solution for $t_{\alpha \beta \omega}$ given in \cite{Osborn:1993cr} satisfying these conditions is:
\begin{equation}
t_{\mu \nu \omega}(X) =   a_1 \frac{X_\mu X_\nu X_\omega}{X^3} + a_2 \left( X_\mu \delta_{\nu \omega}+X_\nu \delta_{\mu \omega} - X_\omega \delta_{\mu \nu} \right)X^{-1}
\end{equation}

Including terms involving $\epsilon_{\mu\nu\omega}$, we find that
\begin{equation}
 t_{\mu \nu \omega}(X) = b \left( - X_\mu X^\lambda \epsilon_{\nu \omega \lambda} X^{-2} + X_\nu X^\lambda \epsilon_{\mu \omega \lambda} X^{-2} + X_\omega X^\lambda \epsilon_{\mu \nu \lambda} X^{-2} \right)
\end{equation}
is another solution to the above equations.

Translating into our notation, the correlation function is:
\begin{eqnarray*}
 && \cor{j^a_\mu(x_1) j^b_\nu(x_2) j^c_\omega(x_3)} \varepsilon_1^\mu \varepsilon_2^\nu \varepsilon_3^\omega  =
\\ &&
\frac{f^{abc}}{|x_{12}||x_{23}||x_{31}|}
\Big( \frac{a_1}{8} Q_1 Q_2 Q_3  + \frac{a_2}{4} \left((P_1)^2Q_1 + (P_2)^2 Q_2 +  (P_3)^2 Q_3   - Q_1 Q_2 Q_3\right)
\\ &&
+ b (Q_1 S_1 + Q_2 S_2 + Q_3 S_3)\Big)
\end{eqnarray*}
Indeed, using the nonlinear constraint (\ref{cubic}), it is clear that the only allowed parity-preserving tensor structures must be linear combinations of $Q_1Q_2Q_3$ and $(P_1)^2Q_1 + (P_2)^2 Q_2 +  (P_3)^2 Q_3$. It is also easy to see that the only allowed parity-violating tensor structure is $Q_1 S_1 + Q_2 S_2 + Q_3 S_3$.

\end{document}